\documentclass[aps,prb,twocolumn,superscriptaddress,raggedbottom,showpacs,floatfix]{revtex4}
\usepackage{epsfig}
\usepackage{amsmath,amsfonts,amssymb,amsthm}
\usepackage{graphicx,latexsym}

\begin{document}
\title{A nonlocal wave-wave interaction among Alfv\'en waves in an intermediate-$\beta$ plasma}
\author{J. S. Zhao}\email{js_zhao@pmo.ac.cn}
\affiliation{Purple Mountain Observatory, Chinese Academy of Sciences 210008,
Nanjing, China.}\affiliation{Graduate School, Chinese Academy of Sciences,
Beijing, China.}
\author{D. J. Wu \footnote{Corresponding author. Electronic address:
\texttt{djwu@pmo.ac.cn}}} \affiliation{Purple Mountain Observatory, Chinese
Academy of Sciences 210008, Nanjing, China.}
\author{J. Y. Lu}\email{lujy@cma.gov.cn}
\affiliation{National Center for Space Weather, China Meteorology
Administration 100081, Beijing, China.}

%J. S. Zhao$^1$ $^3$, D. J. Wu$^1$ $^\ast$, and J. Y. Lu$^2$
%\\
%$^1$ Purple Mountain Observatory, Chinese Academy of Sciences,
%Nanjing 210008, China. (js\_zhao@pmo.ac.cn, djwu@pmo.ac.cn)
%\\
%$^2$ National Center for Space Weather, China Meteorology
%Administration 100081, Beijing, China. (lujy@cma.gov.cn)
%\\
%$^3$ Graduate School, Chinese Academy of Sciences, Beijing, China.
%\\
%$^\ast$ Corresponding author.
%\author{J. S. Zhao$^1$,$^3$, D. J. Wu$^1$,$^\ast$, and J. Y. Lu$^2$}
%\affiliation{$^1$ Purple Mountain Observatory, Chinese Academy of Sciences, Nanjing 210008, China. (js\_zhao@pmo.ac.cn, djwu@pmo.ac.cn)
% \\ $^2$ National Center for Space Weather, China Meteorology Administration 100081, Beijing, China. (lujy@cma.gov.cn)
% \\ $^3$ Graduate School, Chinese Academy of Sciences, Beijing, China.
% \\ $^\ast$ Corresponding author. }
\date{\today}
%\baselineskip 23 pt
%\newline

\begin{abstract}
A nonlocal coupling mechanism to directly transfer the energy from large-scale Magnetohydrodynamic(MHD) Alfv\'en waves
to small-scale kinetic Alfv\'en waves is presented. It is shown that the interaction between a MHD
Alfv\'en wave and a reversely propagating kinetic Alfv\'en wave can generate another kinetic Alfv\'en wave,
and this interaction exists in the plasmas where the thermal to magnetic pressure ratio is larger
than the electron to ion mass ratio. The proposed nonlocal interaction may have a potential application to account for the
observed electron scale kinetic Alfv\'en waves in the solar wind and solar corona plasmas.
\end{abstract}
\maketitle

\renewcommand{\thesection}{\Roman{section}}
\section{Introduction}
\indent The large-scale magnetodydrodynamic(MHD) Alfv\'en wave \cite{al42}, which is generated
by the magnetic stress tensor, is a nondispersive low frequency
wave mode and can be directly derived from MHD equations.
The dispersive Alfv\'en wave arises due to the
finite-frequency and small-scale modifications \cite{ss00}. The
dispersive Alfv\'en wave is specially called as the kinetic Alfv\'en
wave (KAW) when its perpendicular wavelength is close to the order of
the ion gyroradius \cite{ChenHasegawa74}, the ion acoustic
gyroradius, or the electron inertial length \cite{GoertzBoswell79}.
The KAW can accelerate and heat electrons by its parallel electric
field \cite{Wu03a,Wu03b,Chaston06} and interact with ions through
its perpendicular electric field
\cite{VoitenkoGoossens04,WuYang06,WuYang07}. The properties of the
KAW have been demonstrated by many experimental investigations in
space \cite{Louarn94,Wu96,Chaston99,Stasiewicz00} and
laborotary \cite{Gekelman99,Leneman99,Kletzing03,Vincena04}
plasmas. In particular, the KAW can play an important role in the
acceleration of energetic electrons in Earth's aurora
\cite{WuChao03,Lu03,WuChao04,Lu07,Chaston07} and solar flares
\cite{Wu05,Wu07,Fle08} as well as the heating of solar coronal
plasmas \cite{WuFang99,WuFang03,WuFang07,Bingham09,Wang09}.

Extensive works have been done to discuss the local interaction
process of three Alfv\'en waves in which the magnitudes of three wavenumbers are
comparable \cite{gs95,ng96,brodin90,ss95,voitenko98a,zhao10}.
A large amplitude dispersive Alfv\'en wave can bring about
the local nonlinear decay among themselves
\cite{brodin90,ss95,voitenko98a,zhao10}. Local interaction between two counterpropagating
Alfv\'en wave packets can lead the wave
energy to cascade from the energy injection region
to the energy dissipation region in the MHD turbulence
\cite{gs95,ng96,tu84,zhou90,cranmer03}. Through the local cascade, the KAW can be generated
at scales of the order of the ion inertial length or the ion gyroradius in the solar wind turbulence
\cite{leamon99,Schekochihin09,Alexandrova09,Sahraoui09}, however,
it cannot reach electron scales (the electron inertial length or
the electron gyroradius) due to the
large electron Landau damping in these small scales
\cite{Podesta10}.

Recently, some works have found that the nonlinear coupling of
Alfv\'en waves with different scales can occur
\cite{VoitenkoGoossens05,ss05}. Voitenko and Goossens \cite{VoitenkoGoossens05}
presented a nonlocal interaction among three Alfv\'en waves and
showed that a large-scale MHD Alfv\'en wave can decay parametrically
into two small-scale KAWs. Shukla and Stenflo \cite{ss05} studied
the three-wave interaction involving two KAWs and one
field-aligned Alfv\'en wave with the finite-frequency
modification, and they also showed two small-scale KAWs can be
nonlocally excited by this large-scale Alfv\'en wave. Unlike the
local energy cascade in the MHD turbulence, the wave energy can be
directly transferred from large-scale Alfv\'en waves to small-scale
KAWs in these nonlocal interaction processes.

In this paper, we investigate the nonlocal interaction among one MHD
Alfv\'en wave and two KAWs, and we shall
show that the nonlocal coupling, a MHD Alfv\'en wave +
KAW $\rightarrow$ KAW, may play an important role in generating the KAWs with
electron scales \cite{Alexandrova09,Sahraoui09}. The reminder of
this paper is organized in the following fashion. The qualitative
and quantitative analyses are given in section 2 and section 3,
respectively. Section 4 presents an application in the solar
corona. The discussion is set in section 5 and the summary is
contained in section 6.

\section{Qualitative analysis}
Three waves in the nonlinear coupling process (a MHD Alfv\'en wave
+ KAW 1 $\rightarrow$ KAW 2) must satisfy the resonant relation, which
describes the phase relation of these three waves,
\begin{eqnarray}
 \omega_{s}+\omega_{1}=\omega_{2}
 \nonumber\\
 \textbf{k}_{s}+\textbf{k}_{1}=\textbf{k}_{2}
 \label{resonant relation}
\end{eqnarray}
where $\omega_s$ and $\textbf{k}_s$ are the frequency and wave
vector of the large-scale MHD Alfv\'en wave, respectively;
$\omega_{1,2}$ and $\textbf{k}_{1,2}$ are frequencies and wave
vectors of the two KAWs, respectively. In this study, we consider the general
oblique propagating MHD Alfv\'en wave, where wave vector
$\textbf{k}_{s}=\textbf{k}_{s\perp}+k_{sz}\hat{\textbf{z}}$, and
$\textbf{k}_{s\perp}$ and $k_{sz}\hat{\textbf{z}}$ are the wave vectors
perpendicular and parallel to the background magnetic field
$B_0\hat{\textbf{z}}$, respectively.

The linear dispersion relation of the MHD Alfv\'en wave is
$\omega_{s}=V_{A}k_{sz}$, where $V_{A}$ is the Alfv\'en velocity.
For the two KAWs in (\ref{resonant relation}), their linear dispersion
relations are given as $\omega_{1,2}=V_{A}k_{1,2z}K_{1,2}$ with
$K_{1,2}=\sqrt{1+\rho^2k_{1,2\perp}^2}$ for the intermediate-beta
plasmas ($m_e/m_i \ll \beta\ll 1$) and
$K_{1,2}=1/\sqrt{1+\lambda_{e}^2k_{1,2\perp}^2}$ for the low-beta
plasmas ($\beta\ll m_e/m_i$) \cite{ss00,lysak96}, where $k_{1,2z}$ and $k_{1,2\perp}$
are the parallel and perpendicular wavenumbers of the two KAWs, respectively, $\rho^2=\rho_i^2+\rho_s^2$
($\rho_i$ is the ion gyroradius and $\rho_s$ is the ion acoustic
gyroradius), $\lambda_{e}$ is the electron inertial length,
$\beta$ is the ratio of the plasma thermal to magnetic pressures,
and $m_e/m_i$ is the electron to ion mass ratio.

Substituting the linear dispersion relations of the MHD Alfv\'en
wave and two KAWs into the frequency relation in (\ref{resonant
relation}) and then combining the z-direction wave-vector
relation, we can obtain a restricted relation for two
wavenumbers of the two KAWs,
\begin{equation}
 k_{1z}(K_{1}-s_{1})=k_{2z}(K_{2}-s_{2}),
 \label{restrict relation1}
\end{equation}
or be written as
\begin{equation}
 (K_{1}-s_{1})(K_{2}-s_{2})>0.
 \label{restrict relation2}
\end{equation}
where the MHD Alfv\'en wave is assumed along the background magnetic
field. The subscribes $s_{1,2}$ denote the directions
of the two KAWs, for example, $s_{1,2}=1$ stand for two waves having
the same directions as the MHD Alfv\'en wave and $s_{1,2}=-1$ are
two reversely-propagating waves.

There exist three kinds of wave-wave interaction, $s_{1,2}=\pm1$
and $s_{1}=-s_{2}=-1$, for the coupling of one MHD Alfv\'en wave and
two KAWs. In the first two interaction cases ($s_{1,2}=\pm1$), the
frequency of the KAW 1 is required much larger than the MHD
Alfv\'en wave frequency, and we do not discuss these two cases in
this study. The third case ($s_{1}=-s_{2}=-1$), as shown in Fig.
1, describes the parallel-propagating KAW 2 generated in the
nonlinear interaction between the parallel-propagating MHD
Alfv\'en wave and the reversely-propagating KAW 1. From the
restricted relation Eq. (\ref{restrict relation2}), we can see that
the third interaction takes place under the condition
$K_{2}>1$, which means this interaction only exists in the plasmas
with $\beta>m_{e}/m_{i}$. Further, we can obtain
$\omega_{1}=\omega_{s}(K_{2}-1)/(K_{2}/K_{1}+1)$ and
$\omega_{2}=\omega_{s}(K_{1}+1)/(K_{1}/K_{2}+1)$ from Eqs.
(\ref{resonant relation}) and (\ref{restrict relation1}).

\begin{figure}
  \centerline{\includegraphics[width=8cm]{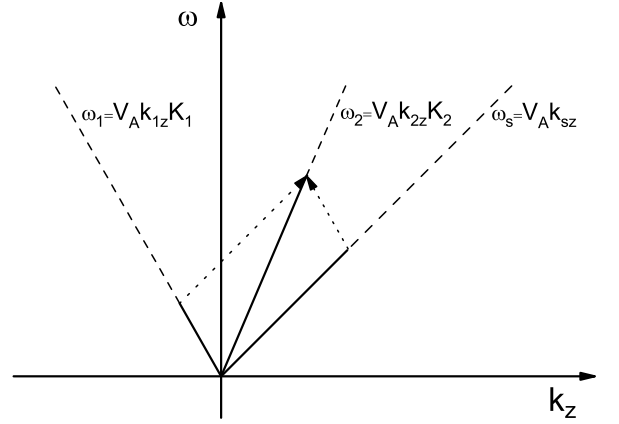}}
  \caption{
    The parallelogram in the ($\omega,k_z$) plane for the three Alfv\'en waves interaction, a MHD
    Alfv\'en wave + KAW 1 $\rightarrow$ KAW 2.
  }
  \label{fig:gamma}
\end{figure}

\section{Two-fluid model}
Let us consider the homogeneous and collisionless plasmas with two
species of particles, electrons and ions. We restrict the
plasmas in the intermediate-beta range ($m_e/m_i \ll \beta \ll 1$), where the ion finite
Larmor radius and ion polarization contributions are
both important for the KAWs \cite{ss00}. The momentum equation is,
\begin{align}
  \partial_{t}\textbf{v}_{j}+&\frac{T_j}{m_{j}n_{j}}\nabla n_{j}-\frac{q_j}{m_{j}}(\textbf{E}
  +\textbf{v}_{j}\times\textbf{B}_{0})= \nonumber\\
  &-\textbf{v}_{j}\cdot\nabla\textbf{v}_{j}+\frac{q_{j}}{m_{j}}\textbf{v}_{j}\times\textbf{B},
  \label{momentum equation}
\end{align}
where $\textbf{v}_{j}$, $m_{j}$, $q_{j}$, $n_{j}$, and $T_j$
denote the particle velocity, mass, charge, number density, and temperature for species $j$, respectively; $\textbf{E}$ and
$\textbf{B}$ are the wave electric and magnetic field perturbations, respectively.
Note that the isothermal assumption has been used in (4).

The variables can be written as,
$\textbf{v}_{j}=\textbf{v}_{js}+\textbf{v}_{j1}+\textbf{v}_{j2}$,
$n_{j}=n_{0}+n_{j1}+n_{j2}$,
$\textbf{E}=\textbf{E}_{s}+\textbf{E}_{1}+\textbf{E}_{2}$ and
$\textbf{B}=\textbf{B}_{s}+\textbf{B}_{1}+\textbf{B}_{2}$ (where
$n_{0}$ is the background number density). The large-scale MHD Alfv\'en wave only has the perpendicular
velocity and electromagnetic perturbations:
$\textbf{v}_s=\textbf{v}_{s\perp}$,
$\textbf{E}_{s}=\textbf{E}_{s\perp}$ and
$\textbf{B}_s=\textbf{B}_{s\perp}$. By substituting these expressions
into the momentum Eq. (\ref{momentum equation}), we can
obtain the fluid velocities of the KAW 2,
\begin{align}
 &\textbf{v}_{i2\perp}\simeq \frac{1}{B_0}\textbf{E}_{2\perp}\times\hat{\textbf{z}}
 -\frac{v_{Ti}^2}{\omega_{ci}}\nabla_{\perp}\frac{n_{i2}}{n_{0}}\times\hat{\textbf{z}}
 +\frac{1}{B_0\omega_{ci}}\partial_{t}\textbf{E}_{2\perp}
  \nonumber\\
 &-\frac{v_{Ti}^2}{\omega_{ci}^2}\partial_{t}\nabla_{\perp}\frac{n_{i2}}{n_{0}}
 -\frac{1}{\omega_{ci}}[\textbf{v}_{is\perp}\cdot\nabla_{\perp}(\textbf{v}_{i1E}+\textbf{v}_{i1D})]\times\hat{\textbf{z}},
\label{ion perpendicular velocity}
\end{align}
\begin{align}
 \textbf{v}_{e2\perp}\simeq &\frac{1}{B_0}\textbf{E}_{2\perp}\times\hat{\textbf{z}}
 -\frac{v_{Te}^2}{\omega_{ce}}\nabla_{\perp}\frac{n_{e2}}{n_{0}}\times\hat{\textbf{z}}
 \nonumber\\
 &+\frac{1}{B_0}(\textbf{v}_{e1z}\times\textbf{B}_{s\perp})\times\hat{\textbf{z}},
 \label{electron perpendicular velocity}
\end{align}
and
\begin{align}
 &\partial_{t}v_{e2z}\simeq-\frac{e}{m_{e}}E_{2z}-v_{Te}^2\partial_{z}\frac{n_{e2}}{n_{0}}
 \nonumber\\
 -\frac{e}{m_{e}}[&(\textbf{v}_{e1E}+\textbf{v}_{e1D})\times\textbf{B}_{s\perp}
 +\textbf{v}_{es\perp}\times\textbf{B}_{1\perp}]\cdot\hat{\textbf{z}}.
 \label{electron parallel velocity}
\end{align}
where $\omega_{ci}=eB_{0}/m_{i}$ and $\omega_{ce}=-eB_{0}/m_{e}$ are the ion and electron cyclotron frequencies,
respectively; $v_{Ti}=(T_i/m_i)^{1/2}$ and $v_{Te}=(T_e/m_e)^{1/2}$ are the ion and electron thermal velocities, respectively;
$\textbf{v}_{i1E}=(1/B_0)\textbf{E}_{1\perp}\times\hat{\textbf{z}}$
and
$\textbf{v}_{i1D}=-(v_{Ti}^2/\omega_{ci})\nabla_{\perp}(n_{i1}/n_{0})\times\hat{\textbf{z}}$
denote the ion electric field drift velocity and the ion
diamagnetic drift velocity of the KAW 1, respectively;
$\textbf{v}_{e1z}$,
$\textbf{v}_{e1E}=(1/B_0)\textbf{E}_{1\perp}\times\hat{\textbf{z}}$
and
$\textbf{v}_{e1D}=-(v_{Te}^2/\omega_{ce})\nabla_{\perp}(n_{e1}/n_{0})\times\hat{\textbf{z}}$
denote the electron parallel velocity, the electric field drift
velocity, and the electron diamagnetic drift velocity of the KAW 1,
respectively. Here, the nonlinear terms due to the thermal
pressure gradient are neglected because they are smaller than the
nonlinear effects in (\ref{ion perpendicular velocity}) -
(\ref{electron parallel velocity}). This
study is limited to the weak nonlinear system, where a Fourier
analysis can be used. That is, the wave amplitudes are restricted as small amplitudes,
and the ratio of energy in the oscillations to the total energy of
the plasma is a small parameter (e.g. \cite{sg69}). The waves in this system are
assumed to be three monochromatic waves, and their amplitudes can be
expressed as $A_l=A_{k_l}e^{-i\omega_l t +i\textbf{k}_l \cdot
\textbf{r}}$, where $l=(1, 2, s)$.

The ion and electron number density perturbations in equations
(\ref{ion perpendicular velocity}) - (\ref{electron parallel
velocity}) can be obtained by using the ion continuity equation
and the quasi-neutrality condition ($n_{2i}=n_{2e}\equiv n_2$),
\begin{eqnarray}
 (1+\rho_{i}^2k_{2\perp}^2)n_2/n_{0}\simeq-\frac{ie}{m_{i}\omega_{ci}^2}\textbf{k}_{2\perp}\cdot\textbf{E}_{2\perp}.
 \label{ion density}
\end{eqnarray}

The wave electromagnetic fields $\textbf{E}_2$ and $\textbf{B}_2$
in (\ref{ion perpendicular velocity}) - (\ref{ion density}) can be
conveniently represented by a scalar potential $\phi_2$ and a
z-direction vector potential $A_{2z}\textbf{z}$,
\begin{eqnarray}
 \textbf{E}_2=-i\textbf{k}_2\phi_2+i\omega_2(A_{2z}\hat{\textbf{z}}),  \  \
 \textbf{B}_2=i\textbf{k}_{2\perp}\times(A_{2z}\hat{\textbf{z}}).
 \label{electromagnetic field}
\end{eqnarray}

By using (\ref{ion perpendicular velocity}) -
(\ref{electromagnetic field}), the current density
$\textbf{J}_2=en_0(\textbf{v}_{i2\perp}-\textbf{v}_{e2\perp}-v_{2ez}\hat{\textbf{z}})$
is written as,
\begin{align}
 &\textbf{k}_{2\perp}\cdot\textbf{J}_{2\perp}\simeq
 -\frac{n_{0}e^2}{m_{i}\omega_{ci}^2}\frac{1}{1+\rho_{i}^2k_{2\perp}^2}\omega_2 k_{2\perp}^2\phi_2
 -\frac{n_{0}e}{\omega_{ci}}\textbf{k}_{2\perp}\cdot
 \nonumber\\
 & \{ [\textbf{v}_{is\perp}\cdot i\textbf{k}_{1\perp}(\textbf{v}_{i1E}+\textbf{v}_{i1D})
 +\frac{e}{m_{i}}\textbf{v}_{e1z}\times\textbf{B}_{s\perp} ]\times\hat{\textbf{z}}\},
\label{j1perp}
\end{align}
and,
\begin{align}
 &\omega_2 J_{2z}\simeq
 \frac{n_{0}e^2}{m_{e}}\frac{1+\rho^2 k_{2\perp}^2}{1+\rho_{i}^2 k_{2\perp}^2}s_2k_{2z}\phi_2
 -\frac{n_{0}e^2}{m_{e}}\omega_2 A_{2z}
 \nonumber\\
 &-\frac{n_{0}e^2}{m_{e}}[(\textbf{v}_{e1E}+\textbf{v}_{e1D})\times\textbf{B}_{s\perp}
 +\textbf{v}_{es\perp}\times\textbf{B}_{1\perp}]\cdot\hat{\textbf{z}},
 \label{j1pal}
\end{align}

The current density can also be given in terms of $A_{2z}$ by
Ampere's law $\mu_{0}\textbf{J}_2=\nabla\times\textbf{B}_2$,
\begin{eqnarray}
\mu_{0}\textbf{J}_{2\perp}=-\textbf{k}_{2\perp} k_{2z}A_{2z}, \ \
\mu_{0}J_{2z}=k_{2\perp}^{2}A_{2z},
 \label{j2}
\end{eqnarray}

From Eqs. (\ref{j1perp}) and (\ref{j2}), a nonlinear equation for
the KAW 2 is given by,
\begin{align}
  &[\omega_{2}^2-V_{A}^2(1+\rho^2 k_{2\perp}^2)k_{2z}^2]\phi_{2}=
 \nonumber\\
 & -B_0(1+\rho_{i}^2k_{2\perp}^2)\omega_{2}k_{2\perp}^{-2}\textbf{k}_{2\perp}\cdot
 \nonumber\\
 \{[\textbf{v}_{is\perp}\cdot i& \textbf{k}_{1\perp}(\textbf{v}_{i1E}+\textbf{v}_{i1D})
 +\frac{e}{m_{i}}\textbf{v}_{e1z}\times\textbf{B}_{s\perp}]\times\hat{\textbf{z}}\}
 \nonumber\\
 &+V_{A}^2(1+\rho_{i}^2k_{2\perp}^2)i s_2k_{2z}
 \nonumber\\
 [(\textbf{v}_{e1E}&+\textbf{v}_{e1D})\times\textbf{B}_{s\perp}
 +\textbf{v}_{es\perp}\times\textbf{B}_{1\perp}]\cdot\hat{\textbf{z}}.
 \label{coupling equation}
\end{align}
where the first nonlinear term on the right-hand of the equation
comes from the ion convective motion and perpendicular electron nonlinear
Lorentz force, and the second nonlinear term arises due
to the parallel electron nonlinear Lorentz force. Three nonlinear
effects (the ion convective motion, the parallel and perpendicular electron
nonlinear Lorentz forces) are all important in the
nonlocal coupling of one MHD Alfv\'en wave and two KAWs,
and these three effects are also contained in some other nonlinear
models in simplified forms (e.g.,
\cite{zhao10,ss05,ss04}).

The variables of the KAW 1 in the nonlinear Eq.
(\ref{coupling equation}) can be expressed in terms of the scalar
potential $\phi_1$,
\begin{align}
 &\textbf{v}_{i1E}=\textbf{v}_{e1E}=-\frac{i}{B_{0}}\textbf{k}_{1\perp}\times\hat{\textbf{z}}\phi_{1},
 \nonumber\\
 &\textbf{v}_{i1D}=\frac{i}{B_{0}}\frac{\rho_{i}^2k_{1\perp}^2}{1+\rho_{i}^2k_{1\perp}^2}\textbf{k}_{1\perp}\times\hat{\textbf{z}}\phi_{1},
 \nonumber\\
 &\textbf{v}_{e1D}=-\frac{i}{B_{0}}\frac{\rho_{s}^2k_{1\perp}^2}{1+\rho_{i}^2k_{1\perp}^2}\textbf{k}_{1\perp}\times\hat{\textbf{z}}\phi_{1},
 \nonumber\\
 &v_{e1z}=-\frac{1}{B_{0}\omega_{ci}}\frac{k_{1\perp}^2}{1+\rho_{i}^2k_{1\perp}^2}\frac{s_{1}\omega_{1}}{k_{1z}}\phi_{1}, \nonumber\\
 &\textbf{B}_{1\perp}=i\frac{s_{1}\omega_{1}}{V_{A}^2k_{1z}}\frac{\textbf{k}_{1\perp}\times\hat{\textbf{z}}}{1+\rho_{i}^2k_{1\perp}^2}\phi_{1},
\label{linear relation KAW 1}
\end{align}

For the obliquely propagating MHD Alfv\'en wave, we use a scalar potential
$\phi_s$ to express the electric field perturbation
$\textbf{E}_{s}=-\nabla_{\perp}\phi_s$ and other variables in Eq.
(\ref{coupling equation}),
\begin{eqnarray}
 \textbf{v}_{is\perp}=\textbf{v}_{es\perp}=-\frac{V_A}{B_0}\textbf{B}_{s\perp}=
 -\frac{i}{B_{0}}\textbf{k}_{s\perp}\times\hat{\textbf{z}}\phi_{s}.\
 \label{linear relation SAW}
\end{eqnarray}

By using linear relations in Eqs. (\ref{linear relation KAW 1}) and
(\ref{linear relation SAW}), we can obtain the nonlinear
dispersion relation of the KAW 2,
\begin{align}
 &(\omega_{2}^2-V_{A}^2k_{2z}^2K_{2}^2)\phi_{2}=
 \frac{i\omega_{2}}{B_{0}}\frac{1+\rho_{i}^2k_{2\perp}^2}{1+\rho_{i}^2k_{1\perp}^2}
 \nonumber\\
 (s_{1}s_{2}\frac{K_{1}}{K_{2}}&+\frac{k_{1\perp}^2}{k_{2\perp}^2})
 (1-s_{1}K_{1})(\textbf{k}_{s\perp}\times\hat{\textbf{z}})\cdot\textbf{k}_{1\perp}\phi_{1}\phi_{s}.
 \label{KAW 2 dispersion relation}
 \end{align}
where $K_{1,2}=\sqrt{1+\rho^2k_{1,2\perp}^2}$ is the definition in
the intermediate-beta plasmas and the relation $k_{1\perp}\simeq
k_{2\perp}\gg k_{s\perp}$ has been used. The nonlinear dispersion
relation of the KAW 1 can be obtained by exchanging the
subscripts 1 and 2 and replacing $\phi_{s}$ by $\phi_{s}^{*}$ in Eq.
(\ref{KAW 2 dispersion relation}). We here restrict the MHD
Alfv\'en wave as the finite amplitude wave and neglect its
nonlinear modification caused by the two KAWs. This finite amplitude
assumption can be permitted for some solar and space plasma
environments, such as the solar corona and the solar wind.

Two KAWs frequencies are expressed as $
\omega_{1}=\omega_{1r}-i\gamma_1$ for the KAW 1 and
$\omega_{2}=\omega_{2r}+i\gamma_2$ for the KAW 2, where
$\omega_{r1,2}$ denote the real parts of two frequencies,
$\gamma_1$ is the decay rate of the KAW 1 and $\gamma_2$ is the
growth rate of the KAW 2. To show which parameters affect the
nonlocal coupling strength among three Alfv\'en waves, we
assume $\gamma_1\simeq \gamma_2\equiv \gamma$ and
$\gamma\ll\omega_{r}$. Then, $\gamma$ can be
derived from the nonlinear dispersion relations of the two KAWs,
\begin{align}
 &\gamma^2=\frac{V_{A}^2}{4}\frac{K_{2}k_{2\perp}^2}{K_{1}}
 \nonumber\\
 (s_{1}s_{2}\frac{K_{1}}{K_{2}}
 +\frac{k_{1\perp}^2}{k_{2\perp}^2})^2&
 (K_{1}-s_{1})(K_{2}-s_{2})\textrm{sin}^2\theta
 \frac{B_{s\perp}^2}{B_{0}^2},
 \label{growth rate}
 \end{align}
where $B_{s\perp}=k_{s\perp}\phi_s/V_A$ and $\theta$ is the angle
between $\textbf{k}_{s\perp}$ and $\textbf{k}_{1\perp}$. From the
equation (\ref{growth rate}), the restrict relation
$(K_{1}-s_{1})(K_{2}-s_{2})>0$ can be reobtained. The equation
(\ref{growth rate}) is rewritten as the following form for the
interaction case $s_1=-s_2=-1$,
\begin{align}
 &\gamma^2=
 \frac{V_{A}^2}{4}\frac{K_{2}k_{2\perp}^2}{K_{1}}\nonumber\\
 (-\frac{K_{1}}{K_{2}}
 +\frac{k_{1\perp}^2}{k_{2\perp}^2})^2&
 (K_{1}+1)(K_{2}-1)\textrm{sin}^2\theta
 \frac{B_{s\perp}^2}{B_{0}^2}.
 \label{growth rate2}
 \end{align}

\section{Application in the solar corona}
Here we use some typical parameters in the quiet solar corona to
discuss the nonlocal interaction proposed in this study: the
number density $n=10^9$ $\textrm{cm}^{-3}$, the thermal temperature
$T_{i}=T_{e}=10^6$ $\textrm{K}$, and the magnetic field $B_{0}=10$
$\textrm{G}$. These values give the plasma beta
$\beta\sim3.5\times10^{-2}$ that is located in the
intermediate-beta range. The existence of the MHD Alfv\'en wave
has been demonstrated in the solar corona plasmas by many
observations \cite{jess09,To09,He09}. The frequencies of these MHD
Alfv\'en waves are in the range ($10^{-5}$ Hz, $10^{-1}$ Hz) \cite{To09,He09,cr05}, or the high frequency range
(0.1 Hz, 2.5 Hz) if these waves are generated
by the reconnection in the magnetic networks \cite{ryutova01}. As
an example, we consider two MHD Alfv\'en
waves with the same frequencies $10^{-1}$ Hz but different
directions, $k_{s\perp}=k_{sz}$ ($\rho k_{s\perp}\sim
10^{-7}$) and $k_{s\perp}=10k_{sz}$ ($\rho k_{s\perp}\sim
10^{-6}$). If the two KAWs locate in the frequency range ($10^{-5}$ Hz, 2.5 Hz), we can use the frequency relation
$\omega_{1}=\omega_{s}(K_{2}-1)/(K_{2}/K_{1}+1)$ to estimate the
limitation of the perpendicular wavenumber for the KAW 1.
It shows that $\rho k_{1\perp}$ is required  larger
than $0.02$, meanwhile, we can also obtain the limitation of $\rho k_{2\perp}$ because of $\rho k_{1\perp}\simeq \rho k_{2\perp}$.
\begin{figure}
  \centerline{\includegraphics[width=8cm]{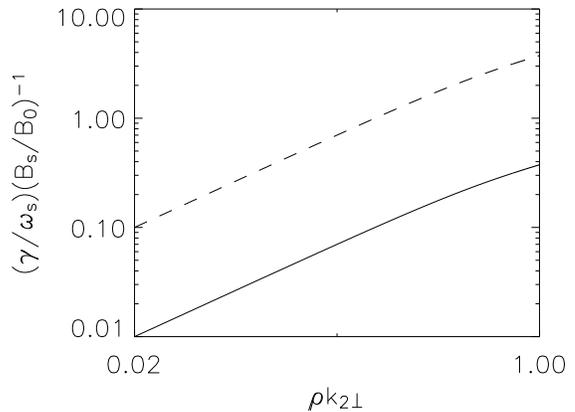}}
  \caption{
    The relation between the maximal coupling strength $\gamma$ and the
    perpendicular wavenumber of the KAW 2 $\rho k_{2\perp}$.
    The solid line denotes the MHD Alfv\'en wave propagating case $k_{s\perp}=k_{sz}$
    and the dashed line is for the case $k_{s\perp}=10k_{sz}$.
  }
  \label{fig:gamma}
\end{figure}

Figure 2 gives the relation between the maximal coupling strength
$\gamma$ and the perpendicular wavenumber of the KAW 2 $\rho
k_{2\perp}$, where the solid line denotes the MHD Alfv\'en wave
propagating case $k_{s\perp}=k_{sz}$ and the dashed line is for
the case $k_{s\perp}=10k_{sz}$. Figure 2 shows that the coupling
strength $\gamma$ increases with the increment of the $\rho
k_{2\perp}$ and the $\gamma$ is large for the more obliquely propagating MHD
Alfv\'en wave.

The new generated KAW can suffer the electron Landau damping in
the collisionless plasmas \cite{GaryBoro08}, and the expression of
the electron Landau damping rate is given as
$\gamma_{L}/\omega_{ci}=A(m_{e}/m_{i})^{1/2}\beta_{e}^{1/2}(\lambda_{i}k_{\perp})^2\lambda_{i}k_{z}
$ \cite{GaryBoro08}, where $\beta_{e}$ is the electron plasma
beta, $\lambda_{i}$ is the ion inertial length and $A$ is a function
of $\beta_{e}$ and $T_{e}/T_{i}$. For the low beta plasmas with
$\beta_{e}\ll 1$, $A$ is roughly 0.4. We can use the electron Landau damping rate and the expression (\ref{growth
rate2}) to give some estimations for the threshold amplitude of
the MHD Alfv\'en wave.

Figure 3 gives the threshold amplitude of the MHD Alfv\'en wave as
a function of the $\rho k_{2\perp}$. Figure 3 shows that the
threshold amplitude is smaller for the more obliquely propagating MHD Alfv\'en
wave and for the KAWs with smaller $\rho k_{2\perp}$. In the solar
low corona, the relative amplitude of the MHD Alfv\'en wave nearly
reaches $0.2$ \cite{He09}, which is larger than all the threshold
amplitudes shown in Figure 3. A small threshold amplitude,
especially for the more obliquely propagating MHD Alfv\'en wave,
implies that the nonlocal interaction mechanism may happen in the
solar corona.
\begin{figure}
  \centerline{\includegraphics[width=8cm]{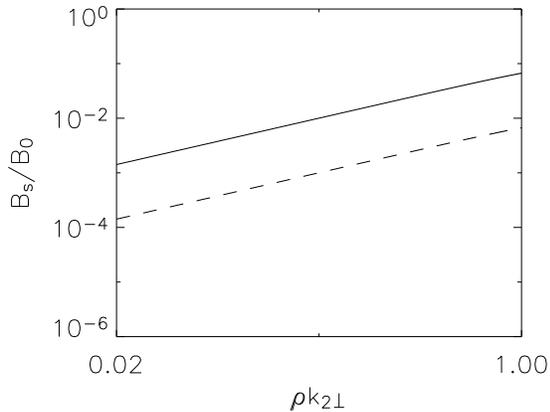}}
  \caption{
    The threshold amplitude of the MHD Alfv\'en wave changes with the
    perpendicular wavenumber of the KAW 2 $\rho k_{2\perp}$. The lines have the same
    meaning as in Fig. 2.
  }
  \label{fig:gamma}
\end{figure}

\section{Discussion}
Our nonlinear wave-wave interaction requires a large-scale MHD Alfv\'en wave and
a small-scale KAW with counterpropagating directions. These two waves can coexist in the solar and
space plasmas. The MHD Alfv\'en wave is a common wave mode and can
be easily generated by the shear motion and/or the magnetic
reconnection process. The KAW can be produced by many mechanisms,
such as the phase mixing and the resonant absorption of the MHD
Alfv\'en waves \cite{HasegawaChen75,Ionson78,HeyvaertsPriest83},
the instability caused by the warm proton beams or by the ion-ion
streaming \cite{voitenko98b,Lin07}, the velocity shear transition
\cite{chen05}, and the parametric decay by other wave modes
\cite{shukla78,voitenko03}. These wave generation mechanisms make
it easily to produce the opposite traveling MHD Alfv\'en wave and KAW, thus
causing the nonlocal interaction.

The mechanism proposed in this study is also different from the
other two kinds of interactions among three Alfv\'en waves, the
local interaction and the nonlocal decay
\cite{gs95,ng96,tu84,zhou90,cranmer03,VoitenkoGoossens05,ss05}.
The proposed wave-wave interaction happens for
Alfv\'en waves with different scales, whereas the local interaction is limited to waves
with comparable scales \cite{gs95,ng96,tu84,zhou90,cranmer03}.
The nonlocal decay mechanism also relates to three waves with
different scales, however, it describes two waves excited by a pump wave
\cite{VoitenkoGoossens05,ss05}. Of course, two nonlocal
interactions, either through the nonlocal decay or through our nonlocal
interaction, can both directly transport the Alfv\'en
wave energy from the large-scale region to the small-scale region
and then dissipate the wave energy there.

Last, let us give a simple discussion for a potential application
of our nonlocal mechanism. Recent observations showed there exists
the KAW even down to the electron scales (the electron
inertial length or the electron gyroradius scale) in the solar
wind \cite{Alexandrova09,Sahraoui09,Sahraoui10}. This electron
scale KAW cannot be explained by the local wave cascade
mechanism because the large electron Landau damping can totally
dissipate the KAW energy in that small scale region
\cite{Podesta10}. But our results implies, except for the local
wave-wave interaction, the nonlocal wave-wave interaction may also
happen in the Alfv\'en wave turbulence. Furthermore, as shown in
figure 2, the nonlocal coupling is stronger for the smaller
scale KAWs. So the observed electron scale KAWs may be produced by
this nonlocal mechanism. It should be pointed out that the nonlocal
turbulence of the KAWs still receives little attention and this
problem will be further investigated in our future works.

\section{Conclusion}
This paper discusses the nonlocal coupling of one MHD
Alfv\'en wave and two KAWs, a MHD Alfv\'en wave + KAW 1 $\rightarrow$ KAW 2. The qualitative
discussion shows that this wave-wave interaction works in the
plasmas with $\beta>m_{e}/m_{i}$. The frequency relations among
the two KAWs and the MHD Alfv\'en wave are
$\omega_{1}=\omega_{s}(K_{2}-1)/(K_{2}/K_{1}+1)$ and
$\omega_{2}=\omega_{s}(K_{1}+1)/(K_{1}/K_{2}+1)$. The quantitative
discussion shows that the coupling is stronger for the more
obliquely propagating MHD Alfv\'en wave and for the KAWs with lager
perpendicular wavenumbers. We also show that the proposed mechanism
may play an important role in generating electron scale KAWs in the solar wind.

\section{Acknowledgment}
The authors acknowledge referees for their assistances in
evaluating the manuscript. This study has been supported by the
NSFC grants 10973043, 40874087, 41074107 and 41031063, the NKBRSF under 2011CB81102, the CMA grant GYHY201106011,
and the CAS Special
Grant for Postgraduate Research, Innovation and Practice.
Financial support by Ocean Public Welfare Scientific Research
Project, State Oceanic Administration People's Republic of China
(No.201005017) is also acknowledged.

\end{document}